\documentclass[showpacs,aps,rmp,twocolumn]{revtex4-1}
\usepackage{amsmath}
\usepackage{amsfonts}
\usepackage{amssymb}
\usepackage{amsthm}
\usepackage{graphicx}
\usepackage{color}
\begin{document}

%----------------------------------------------------------------%

\title{Persistent Spin Textures in Semiconductor Nanostructures}
\author{John Schliemann}
%\email[To whom correspondence should be addressed. Electronic
%address: ]{John.Schlliemann@physik.uni-regensburg.de}
\affiliation{ Institute for Theoretical Physics, University of
Regensburg, D-93040 Regensburg, Germany}
\date{July 2016}

\begin{abstract}
Device concepts in semiconductor spintronics make long spin lifetimes
desirable, and the requirements put on spin control by
schemes of quantum information processing are even more demanding. 
Unfortunately, due to spin-orbit coupling
electron spins in semiconductors are generically subject to rather fast
decoherence. In two-dimensional quantum wells made of zinc-blende 
semiconductors, however, the spin-orbit
interaction can be engineered in such a way that persistent spin structures with
extraordinarily long spin lifetimes arise even in the presence of disorder 
and imperfections. We review experimental and theoretical developments
on this subject both for $n$-doped and $p$-doped structures, and we discuss
possible device applications.
\end{abstract}
%\keywords{}

% 71.70.Ej Spin-orbit coupling, Zeeman and Stark splitting, Jahn-Teller effect
% 85.75.-d Magnetoelectronics; spintronics: devices exploiting spin polarized 
%          transport or integrated magnetic fields
% 85.75.Hh Spin polarized field effect transistors

\maketitle

\tableofcontents

\section{Introduction}
\label{intro}

The field of semiconductor spintronics emerged around the turn of the
millennium and comprises a broad variety of efforts towards utilizing the spin 
degree of freedom of electrons, instead, or combined with, their 
charge for information processing, or, even more ambitious, for 
quantum information processing \cite{Zutic04,Fabian07,Wu10}. 
Most activities in this
area rely on the relativistic effect of spin-orbit coupling
described by the Dirac equation and its nonrelativistic
expansion in powers of the inverse speed of light $c$. 
The well-known spin-orbit coupling term arises here in second order,
\begin{equation}
{\cal H}_{so}=\frac{\hbar}{4m_0c^2}\vec\sigma\cdot
\left(\nabla V\times\frac{\vec p}{m_0}\right)\,,
\label{sogeneral}
\end{equation}
where the Pauli matrices $\vec\sigma$ describe the electron\rq s spin,
$m_{0}$ and $\vec p$ are its bare mass and momentum, respectively, 
and $V$ is the potential acting on the particle. 
Moreover, the free Dirac equation, $V=0$, has two dispersion branches with
positive and negative energy, $\varepsilon(\vec p)=\pm\sqrt{m_0^2c^4+c^2p^2}$,
separated by a gap of $2m_0c^2\approx 1{\rm MeV}$,
and the nonrelativistic expansion of the Dirac equation can be viewed
as a method of systematically including the effects of the negative-energy
solutions on the states of positive energy starting from their nonrelativistic
limit. Importantly, the large energy gap $2m_{0}c^{2}$ appears in the
denominator of the right hand side of Eq.~(\ref{sogeneral}) and thus suppresses
spin-orbit coupling for weakly bound electrons.

\begin{figure}
\includegraphics[width=\columnwidth]{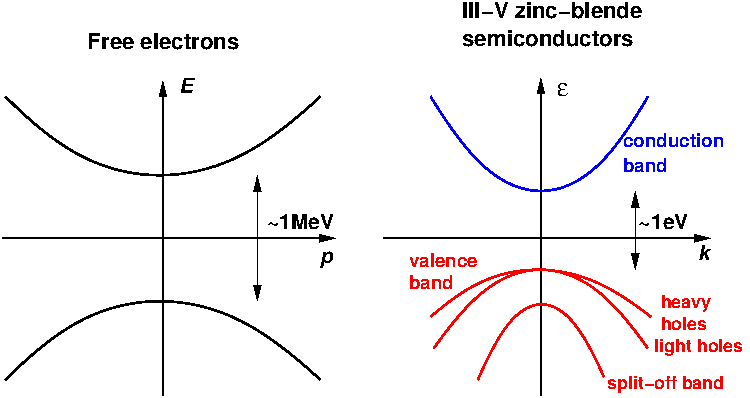}
\vspace{0.2cm}
\caption{(Color online)
Left: Dispersion relation of free electrons showing a gap of about
1 MeV between solutions of positive and negative energy. Right: Schematic
band structure of III-V zinc-blende semiconductors with a band gap of 
typically 1 eV. The $p$-type valence band consists of the heavy and light
hole branches, and the split-off band.}
\label{bandstructure}
\end{figure}

Turning to semiconductors, 
the band structure of zinc-blende III-V systems
exhibits many formal similarities to the situation of free 
relativistic electrons (as sketched in Fig.~\ref{bandstructure}),
while the relevant energy scales are grossly different
\cite{Yu10}. For not too large doping, 
only the band structure around the $\Gamma$ point matters consisting of
a parabolic $s$-type conduction band and a
$p$-type valence band with dispersion branches
for heavy and light holes, and the split-off band. However, the 
fundamental gap between 
conduction and valence band is of order $1{\rm eV}$ or smaller. This heuristic
argument makes plausible that spin-orbit coupling is a significant effect in
III-V semiconductors and actually lies at the very heart of
spintronics research.
\begin{figure}
\includegraphics[width=\columnwidth]{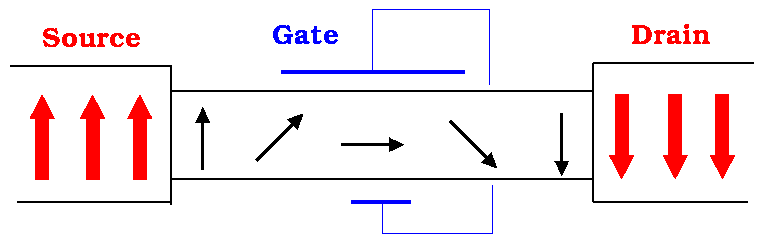}
\vspace{0.2cm}
\caption{(Color online)
Schematic of a spin field-effect transistor: An electron is emitted from a 
spin-polarized source and enters a semiconductor region with spin-orbit
coupling being externally controllable via a perpendicular gate volatge.
In the aboove example the spin-orbit interaction reverses the electron\rq s
spin during its path, and since the drain electrode is polarized opposite
to the source, the conductance of the device is high.}
\label{spinFET}
\end{figure}

A paradigmatic example of a semiconductor spintronics device is the spin
field-effect transistor \cite{Datta90} schematically depicted in 
Fig.~\ref{spinFET}. In this device proposal, an electron
enters a semiconductor region where its spin is rotated via externally
manipulable spin-orbit interaction in a controlled way
such that the carrier is then transmitted into or, depending on the
spin state, reflected from a spin-polarized detector electrode.
A shortcoming of this concept is that impurities and other imperfections act
as scatterers which change the momentum of the electron (i.e. an orbital
degree of freedom) and therefore, again via spin-orbit coupling, easily
also randomize the spin, a process known as the Dyakonov-Perel mechanism
of spin dephasing \cite{Dyakonov72}. 
A way to circumvent this effect is to engineer
the total spin-orbit field acting on the electron spin in such a way that
additional symmetries and related conserved quantities arise which lead
to persistent spin structures. This concept has developed many theoretical
ramifications and manifested itself in various transport and spectroscopic
experiments. These theoretical possibilities and experimental achievements 
are reviewed in the present paper.

This article is organized as follows: Section \ref{ndopedqw}
deals with persistent spin structures in $n$-doped III-V zinc-blende
semiconductor quantum wells and is the main body of this review. In section
\ref{soc-growthdir} we introduce the contributions to spin-orbit coupling
for quantum wells grown along the high-symmetry directions of the
crystal. Section \ref{pshbasic} provides a self-contained discussion
of the theoretical foundation of conserved spin quantities in [001]
quantum wells, but treats also the other high-symmetry growth directions.
The semiclassical description of spin densities
via diffusion equations and their relation to the persistent spin helix
are covered in section \ref{sdiff}. In section \ref{expsim} we report on
the plethora of experiments investigating these predictions along with
pertaining further theoretical work. Theoretical results regarding
signatures of the persistent spin helix arising from many-body physics
are summarized in section \ref{manybody}. Section \ref{ogrowgeo}
is devoted to $n$-doped systems of other geometries including
quantum wells of different growth directions and curved structures.
In section \ref{appl} we summarize developments regarding spin-field-effect
transistors and persistent spin textures. 
Section \ref{omat} contains a discussion of similar
persistent spin structures predicted to occur in materials other than
$n$-doped zinc-blende semiconductors. We close with an outlook in
section \ref{concl}

\section{$n$-Doped Quantum Wells}
\label{ndopedqw}

\subsection{Spin-Orbit Coupling and Growth Direction}
\label{soc-growthdir}

We now summarize important features of the effective description of spin-orbit
interaction in zinc-blende
III-V semiconductors such as GaAs, InAs etc. focusing on two-dimensional
quantum wells \cite{Yu10,Winkler03,Fabian07,Korn10}. 
As already mentioned, due to the lower carrier densities in such
systems compared to, e.g., metals, we can concentrate on the vicinity
of the $\Gamma$-point, i.e. on wave vectors being small compared to the
inverse lattice spacing.

Moreover, we will concentrate here on quantum wells grown into the
high-symmetry directions [001], [110], and [111] which have been in the
focus of theoretical and experimental studies so far. However, very recent
work by \textcite{Kammermeier16} extended the concepts to be discussed below
to more general growth directions.

An important contribution to the effective band structure of 
three-dimensional bulk systems is the Dresselhaus term
given by \cite{Dresselhaus55}
\begin{eqnarray}
{\cal H}_D^{bulk} & = & \gamma\Bigl( 
\sigma^xk_x\left(k_y^2-k_z^2\right)
+\sigma^yk_y\left(k_z^2-k_x^2\right)\nonumber\\
 & & \quad+\sigma^zk_z\left(k_x^2-k_y^2\right)\Bigr)
\label{bulkdressel}
\end{eqnarray}
with the electron\rq s (Bloch) wave vector $\vec k$ and a material parameter 
$\gamma$. This contribution is symmetry-allowed, $\gamma\neq 0$,
due to {\em bulk-inversion asymmetry},
i.e. the fact that the zinc-blende lattice lacks an inversion center.

In sufficiently narrow quantum wells a simplification
occurs as one can, at low enough temperatures, 
approximate the wave vector components along the growth direction by their 
average within the lowest subband. For a symmetric well grown along the 
crystallographic [001] direction we have $\langle k_z\rangle=0$,
and introducing polar coordinates $\vec k=k(\cos\varphi,\sin\varphi)$
for the in-plane components it follows \cite{Dyakonov86,Iordanskii94}
\begin{eqnarray}
{\cal H}_{D}^{001} & = & 
\beta k\left(\sigma^y\sin\varphi-\sigma^x\cos\varphi\right)\nonumber\\
& & -\beta_3k\left(\sigma^x\cos(3\varphi)+\sigma^y\sin(3\varphi)\right)
\label{dressel001-1}
\end{eqnarray}
with $\beta=\beta_1-\beta_3$ and
\begin{equation}
\beta_1=\gamma\langle k_z^2\rangle\qquad,\qquad
\beta_3=\gamma\frac{k^2}{4}\,.
\label{beta}
\end{equation}
Here as before the $x$- and $y$-direction coincide with 
[100] and [010], respectively.
The higher angular harmonics in the second line of Eq.~(\ref{dressel001-1}) are
cubic in $k$. Neglecting these terms leads to
\begin{equation}
\bar{\cal H}_{D}^{001}= 
\beta\left(k_y\sigma^y-k_x\sigma^x\right)
\label{dressel001-2}
\end{equation}
which contains a contribution strictly linear in wave vector ($\propto\beta_1$)
and and a cubic term ($\propto\beta_3$). The latter is usually a small
correction: To give a practical example, for a rectangular well of width
$L=10{\rm nm}$ we have $\langle k_z^2\rangle=(\pi/L)^2\approx 0.1{\rm nm}^{-2}$.
Assuming now a comparatively large density of 
$n=k_f^2/(2\pi)=5\cdot10^{11}{\rm cm}^{-2}$ with a Fermi wave vector
of $k_f=0.17{\rm nm}^{-1}$ (neglecting spin splitting) one finds
$k_f^2/4=0.007{\rm nm}^{-2}$, i.e. $\beta_3(k_f)/ \beta_1=0.07$.
However, there are reports where the quadratic contribution $\beta_3$ to
the Dresselhaus coefficient $\beta$ was found to be essential in order to
accurately describe experimental data \cite{Walser12,Dettwiler14}.
For simplicity we will refer to the Hamiltonian (\ref{dressel001-2}) as the
{\em linear Dresselhaus term} although it contains cubic corrections.

Quantum wells with other growth directions can be similarly described 
by appropriately rotating wave vector and spin in Eq.~(\ref{bulkdressel}). We
restrict the discussion here on the other high-symmetry directions of the 
cubic lattice \cite{Dyakonov86,Eppenga88}.
For the [110] direction one finds
\begin{equation}
{\cal H}_{D}^{110}=\frac{\beta}{2}k_y\sigma^z
+\frac{3\beta_3}{2}k\sigma^z\sin(3\varphi)
\label{dressel110}
\end{equation}
where the $x$- and $y$-direction are along [00$\bar 1$] and [$\bar 1$10], 
respectively.
The coefficient $\beta$ in the first term is again given
by Eqs.~(\ref{beta}) as $\beta=\beta_1-\beta_3$ and summarizes as above the
$k$-linear contribution and the correction provided by the first-harmonic
part of the cubic contributions, whereas the second term contains the
third-harmonic part. 
Remarkably, both terms couple only to the spin projection in the $z$- 
(or [110]-)direction. 

The Dresselhaus term for the [111] direction reads
\begin{eqnarray}
{\cal H}_{D}^{111} & = & 
\frac{2\beta}{\sqrt{3}}\left(k_y\sigma^x-k_x\sigma^y\right)\nonumber\\
& & +\frac{4\beta_3}{\sqrt{6}}k\sigma^z\sin(3\varphi)
\label{dressel111}
\end{eqnarray}
with the $x$- and $y$-direction pointing along [11$\bar 2$] and [$\bar 1$10].
Here the same comments apply as to Eqs.~(\ref{dressel001-1}) and 
(\ref{dressel110}): The first term describes the $k$-linear part with
cubic correction while the second term contains
the higher angular-harmonic part of the cubic contributions.
Neglecting the third angular-harmonic contributions in 
Eqs.~(\ref{dressel110}) and (\ref{dressel111}) leads again to linear
Dresselhaus terms incorporating cubic corrections in their parameter $\beta$.

The second important ingredient to the effective spin-orbit coupling
in quantum wells is known as the Rashba term and is
due to {\em structure-inversion asymmetry}, i.e. it
occurs for confining potentials failing to be invariant under spatial inversion
along the growth direction \cite{Rashba60,Bychkov84a}. This contribution
is described by the expression
\begin{equation}
{\cal H}_{R}=\alpha\left(k_{x}\sigma^{y}-k_{y}\sigma^{x}\right),
\label{rashba}
\end{equation}
where  the Rashba coefficient $\alpha$ is essentially proportional to
the potential gradient across the quantum well and can therefore be varied 
experimentally. This contribution to spin-orbit interaction is the essential
ingredience to the proposal for a spin field -effect transistor
due to \textcite{Datta90} already mentioned in section \ref{intro}.
The linear Rashba term (\ref{rashba}) is independent of the
growth direction and invariant under rotations in the $xy$-plane of the
quantum well.
Remarkably, the above Hamiltonian has the same functional form
as the $k$-linear Dresselhaus term in Eq.~(\ref{dressel111}) 
for the [111] growth direction. 

Although Rashba coupling was first investigated in semiconductors
 \cite{Rashba60,Bychkov84a}, it is nowadays discussed and studied in a much
wider variety of structures lacking inversion symmetry; for a recent
overview see \textcite{Manchon15}. A further source of spin-orbit coupling
in two-dimensional structures are assymetric interfaces \cite{Fabian07};
such contributions will not be considered in the following.

We note that the Rashba term (\ref{rashba})
can somewhat naively be obtained from the general expression
(\ref{sogeneral}) by inserting a linear potential along the $z$-axis.
This approach, however leads to values for $\alpha$ being several orders
smaller than those inferred from experiments, and a realistic description
has to take into account the influence of other bands in addition to
the conduction band \cite{Darnhofer93,AndradaeSilva97,Winkler03,Fabian07,Wu10}. 
This procedure
is effectively similar to the Foldy-Wouthuysen transformation used in
relativistic quantum mechanics to reduce the full Dirac equation for
four-component spinors to an effective description of the ``conduction band''
comprised by solutions of positive energy \cite{Bjorken65}. 
Here the perturbative treatment
of the negative-energy states (``valence band'') leads to the spin-orbit
coupling term (\ref{sogeneral}), apart from other relativistic 
corrections.

\subsection{Persistent Spin Helix: Basic theory}
\label{pshbasic}
 
Let us first consider quantum wells grown in the [001]-direction.
As a result of a large body of experimental as well as theoretical work
\cite{Fabian07,Wu10},
both the parameters $\beta$ and $\alpha$ lie for typical materials and
growth geometries in the ballpark of about $1.0\dots 100{\rm meV}\AA$. In
particular, the Rashba parameter can be tuned to be equal in magnitude to
the Dresselhaus coefficient, $\alpha=\pm\beta$. As we shall see shortly below, 
this situation gives rise to a prime example of a persistent spin texture in a 
semiconductor nanostructure.
\begin{figure}
\includegraphics[width=\columnwidth]{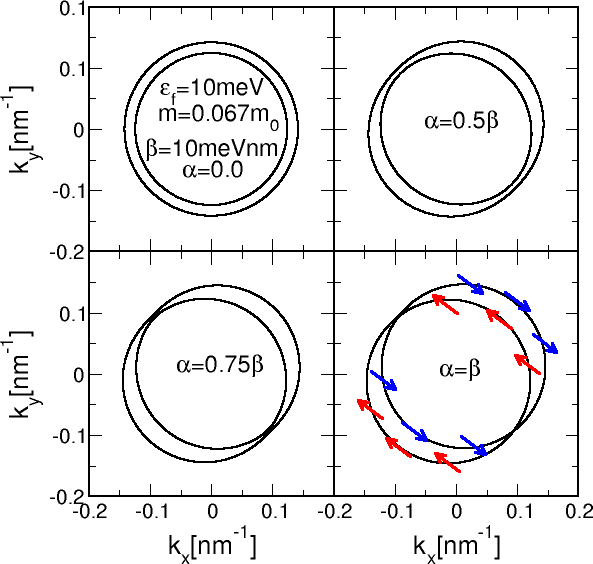}
\vspace{0.2cm}
\caption{(Color online)
Fermi contours for an electron system with band mass
$m=0.067m_0$ (corresponding to GaAs), a typical Fermi energy
of $\varepsilon_f=10{\rm meV}$, and a Dresselhaus parameter of
$\beta=10{\rm meVnm}$. With growing Rashba parameter the energy 
dispersion becomes increasingly anisotropic. For the case $\alpha=\beta$
(bottom right) the spin directions being independent of wave vector are 
indicated. 
Figure adapted from \textcite{Schliemann03b}.}
\label{fig2}
\end{figure}

Let us consider a Hamiltonian consisting of the usual quadratic kinetic energy
characterized by a band mass $m$, and the Rashba (\ref{rashba}) and the
linear Dresselhaus term (\ref{dressel001-2}),
\begin{equation}
{\cal H}=\frac{\hbar^2k^2}{2m}+{\cal H}_R+\bar{\cal H}^{001}_D\,.
\label{ham1}
\end{equation}
leading to the two dispersion branches
\begin{equation}
\varepsilon_{\pm}\left(\vec k\right)
=\frac{\hbar^2k^2}{2m}
\pm k\sqrt{\alpha^2+\beta^2+2\alpha\beta\sin(2\varphi)}
\label{dispersion1}
\end{equation}
which are illustrated in Fig.~\ref{fig2} for different typical 
parameters. As seen from the figure and the above equation, 
the dispersion is clearly anisotropic for $\alpha\neq0\neq\beta$.
Remarkably, this anisotropy in the dispersion relation does not lead to
an anisotropy of the linear electrical bulk conductivity
\cite{Trushin07b,Chalaev08,Chalaev09}, despite an earlier statement in the 
literature \cite{Schliemann03b}.

The case $\alpha=\pm\beta$ shown in the lower right panel of Fig.~\ref{fig2}
is particular \cite{Schliemann03a}: 
Here the Hamiltonian (\ref{ham1}) can be formulated as
\begin{equation}
{\cal H}=\frac{\hbar^2}{2m}
\left(k^2+2\left(\vec k\cdot\vec Q\right)\Sigma\right)
\label{ham2}
\end{equation}
with
\begin{equation}
\Sigma=\frac{\mp\sigma^x+\sigma^y}{\sqrt{2}}\qquad,\qquad
\vec Q=\frac{\sqrt{2}m\alpha}{\hbar^2}(1,\pm 1)
\label{sigmaq}
\end{equation}
such that the spin operator $\Sigma$ is a conserved quantity,
\begin{equation}
\left[{\cal H}, \Sigma\right]=0\,.
\label{conserved1}
\end{equation}
The energy dispersions
\begin{equation}
\varepsilon_{\pm}(\vec k)=\frac{\hbar^2}{2m}
\left(k^2\pm 2\left(\vec k\cdot\vec Q\right)\right)
\label{dispersion2}
\end{equation}
form circles whose centers are displaced from the $\Gamma$ point
by $\mp\vec Q$. Differently from Eq.~(\ref{dispersion1}) the double sign
here refers to the spin eigenvalues determined by
$\Sigma\chi_{\pm}=\pm\chi_{\pm}$ where the eigenspinors read for
$\alpha=+\beta$ 
\begin{equation}
\chi_{\pm}=\frac{1}{\sqrt{2}}\left(
\begin{array}{c}
1 \\
\mp e^{-i\pi/4}
\end{array}
\right)\,,
\label{eigenspinor}
\end{equation}
and for $\alpha=-\beta$ the lower spin component acquires an additional
factor of $(-i)$. In particular, the spin state is independent of the
wave vector, i.e. spin and orbital degrees of freedom are disentangled, and
the Kramers degeneracy enforced by time reversal symmetry
is manifested as 
\begin{equation}
\varepsilon_+(\vec k-\vec Q)=\varepsilon_-(\vec k+\vec Q)\,.
\label{degrel}
\end{equation}

The conservation of the spin component $\Sigma$ expressed in 
Eq.~(\ref{conserved1}) obviously remains intact if a spin-independent
single-particle potential or spin-independent interaction among
the electrons are added to the Hamiltonian. In such a case the 
single-particle wave vector $\vec k$
will in general not be conserved any more and is to be replaced by
a proper momentum operator, $\vec k\mapsto -i\nabla$.
For example, adding an arbitrary scalar potential $V(\vec r)$ to the
Hamiltonian (\ref{ham2}) and inserting the ansatz
\begin{equation}
\psi_{\pm}(\vec r)=e^{-i\vec Q\vec r\Sigma}\chi_{\pm}\phi(\vec r)
=e^{\mp i\vec Q\vec r}\chi_{\pm}\phi(\vec r)
\label{ansatz}
\end{equation}
into ${\cal H}\psi_{\pm}=\varepsilon\psi_{\pm}$ leads to the spin-independent
Schr\"odinger equation
\begin{equation}
\left(-\frac{\hbar^2}{2m}\nabla^2+V(\vec r)\right)\phi(\vec r)
=\left(\varepsilon+\frac{2m\alpha^2}{\hbar^2}\right)\phi(\vec r)\,,
\label{schrodingereq}
\end{equation}
where the energy is shifted by 
$\hbar^2Q^2/(2m)=2m\alpha^2/ \hbar^2$. An analogous
 many-particle Schr\"odinger equation is obtained when adding to
(\ref{ham2}) an arbitrary spin-independent interaction among particles; here
the spin component $\Sigma$ of each electron is separately conserved.

Moreover, comparing the spin state of a general wave function composed
of the states (\ref{ansatz}) at given energy,
\begin{eqnarray}
\psi(\vec r) & = & \nu_+\psi_+(\vec r)+\nu_-\psi_-(\vec r)
\nonumber\\
& = & \left(\nu_+e^{-i\vec Q\vec r\Sigma}\chi_+
+\nu_-e^{-i\vec Q\vec r\Sigma}\chi_-\right)\phi(\vec r)
\label{genstate}
\end{eqnarray}
at two arbitrary locations, say $\vec r=0$ and $\vec r=\vec a$, we
see that the spin state of $\psi(\vec a)$ emerges from $\psi(0)$ by
applying the operator $\exp(-i\vec Q\vec a\Sigma)$. This is a 
{\em controlled rotation} being independent of any further detail of the
system encoded in the single-particle potential or the interaction.
As the rotation operator is also independent of energy, this
observation also holds for arbitrary linear combinations of states of different
energy. Thus, under these very general circumstances, the electron spin
undergoes a controlled rotation as a function of position, a phenomenon
dubbed later on the {\em persistent spin helix} \cite{Bernevig06}.

The angle of the above controlled rotation is $2\vec Q\vec a$
and naturally depends on the distance $\vec a$,
while the rotation axis in spin space is defined by 
the conserved operator $\Sigma$
and given by $(\mp 1,1,0)$, depending on $\alpha=\pm\beta$. As a consequence,
the spin component in this direction is constant 
as a function of both position and
time leading to an {\em infinite spin lifetime} as measured by
expectation value of $\Sigma$. The latter feature is of course just a
general property of any conserved operator within an equilibrium state.

Prior to the work by \textcite{Schliemann03a} also other authors reported
peculiarities of the system (\ref{ham1})
at $\alpha=\beta$ although not relating their observations to
the existence of a new conserved quantity. \textcite{Kiselev00a} have
studied an effective spin model of the type (\ref{ham1}) where the
momentum $\vec p=m\dot{\vec r}$ is a classical quantity (also neglecting the 
difference between canonical and kinetic momenta due to spin-orbit coupling)
whose time dependence is generated by a Markov chain modeling independent
elastic scattering events. Here for general Rashba and Dresselhaus parameters,
the spin is rapidly randomized via the Dyakonov-Perel spin relaxation
mechanism \cite{Dyakonov72}.
For $\alpha=\pm\beta$ the time ordering ${\cal T}$
in the time evolution operator 
${\cal U}(t)={\cal T}\exp(-i\int_0^tdt^{\prime}{\cal H}(p(t^{\prime})/ \hbar))$ 
becomes trivial such that it takes, up to a global phase factor, the form
of the above global spin rotation operator \cite{Schliemann03a},
\begin{equation}
{\cal U}(t)=\exp\left(-i\vec Q\vec a(t)\Sigma\right)
\label{timeev}
\end{equation}
with $\vec a(t)=\vec r(t)-r(0)$. In particular \textcite{Kiselev00a}
observed a diverging spin lifetime for expectation values of $\Sigma$;
a similar conclusion was reached by \textcite{Cartoixa03}, slightly subsequent
to the work by \textcite{Schliemann03a}. The suppressed relaxation of
appropriate spin components was also found earlier by \textcite{Averkiev99}.

In another theoretical investigation \textcite{Pikus95} concluded that
contributions of Rashba and Dresselhaus spin-orbit coupling to the electrical 
conductivity cancel each other at $\alpha=\beta$ (albeit both terms were 
predicted there to contribute additively to spin relaxation, in contrast to the
results demonstrated above). \textcite{Tarasenko02} studied the combined
influence of Rashba and Dresselhaus contributions to the beating patterns of 
Shubnikov-de Haas oscillations and predicted an effective cancellation
of the terms at $\alpha=\beta$. Finally, the very fact that Rashba and
Dresselhaus spin-orbit coupling can nontrivially interfere rather than
simply add up was already observed theoretically by \textcite{Knap96}
within investigations of weak localization phenomena.

Writing the Hamiltonian (\ref{ham2}) in the form
\begin{equation}
{\cal H}=\frac{1}{2m}\left(\vec p+\hbar\vec Q\Sigma\right)^2
-\frac{\hbar^2\vec Q^2}{2m}
\label{ham3}
\end{equation}
suggests to interpret the operator $\exp(-i\vec Q\vec r\Sigma)$ occurring
in Eqs.~(\ref{ansatz}),(\ref{genstate}) as a gauge transformation and
$\vec p$ as a gauge-dependent canonical momentum
\cite{Chen08,Tokatly10}. Moreover, since the
hermitian $2\times 2$-matrix $\Sigma$ generates SU(2)-transformations, the
question arises whether the Hamiltonian (\ref{ham3}) admits further
symmetries furnishing a full representation of the Lie algebra su(2).
Following \textcite{Bernevig06} this can be achieved by writing the Hamiltonian
in second-quantized form,
\begin{equation}
{\cal H}=\sum_{\vec k\eta}\frac{\hbar^2}{2m}
\left(k^2+2\eta\vec Q\vec k\right)c^+_{\vec k\eta}c_{\vec k\eta}
\label{ham4}
\end{equation}
along with 
\begin{equation}
T^3=\frac{\Sigma}{2}=\sum_{\vec k\eta}\frac{\eta}{2}c^+_{\vec k\eta}c_{\vec k\eta}
\label{T3}
\end{equation}
where $c^+_{\vec k\eta}$ ($c_{\vec k\eta}$) creates (annihilates) an electron
with wave vector $\vec k$ and spin state $\chi_{\eta}$, $\eta=\pm$.
Defining now
\begin{equation}
T^+_{\vec Q}=\sum_{\vec k}c^+_{(\vec k-\vec Q)+}c_{(\vec k+\vec Q)-}
\label{Tplus}
\end{equation}
and its adjoint $T^-_{\vec Q}=(T^+_{\vec Q})^+$ one easily verifies that
the latter two operators fulfill together with $T^3$ the su(2)
commutation relations,
\begin{equation}
\left[T^+_{\vec Q},T^-_{\vec Q}\right]=2T^3
\quad,\quad\left[T^3,T^{\pm}_{\vec Q}\right]=\pm T^{\pm}_{\vec Q}
\label{su2T}
\end{equation}
and commute, just as $T^3$, with the Hamiltonian, 
\begin{equation}
\left[{\cal H},T^{\pm}_{\vec Q}\right]=0\,.
\label{commT}
\end{equation}
Moreover, since $T^{\pm}_{\vec Q}$ also commute with any Fourier component
of the density $\rho_{\vec q}=\sum_{\vec k\eta}c^+_{\vec k\eta}c_{(\vec k+\vec q)\eta}$,
\begin{equation}
\left[\rho_{\vec q},T^{\pm}_{\vec Q}\right]=0\,,
\label{commrho}
\end{equation}
Eq.~(\ref{commT}) remains also valid if arbitrary spin-independent potentials
or interactions are added to the Hamiltonian. We note that the su(2) commutation
relation (\ref{su2T}) as well as the property (\ref{commrho}) hold for
any vector $\vec Q$. The vanishing of the commutator (\ref{commT}), however,
depends on the specific form given in Eq.~(\ref{sigmaq}) and the fact
that the spin-independent part of the kinetic Hamiltonian (\ref{ham2})
is quadratic in the wave vector leading to the degeneracy (\ref{degrel}). 
For instance, if a term quartic in the
momentum were present (still consistent with time reversal symmetry), 
Eq.~(\ref{commT}) would not hold, and also a
formulation (\ref{ham3}) is the style of a gauge theory would not be
possible.

Applying an in-plane magnetic field perpendicular to $\vec Q$
(i.e. in the direction defined by $\Sigma$) changes the Hamiltonian as
\begin{equation}
{\cal H}^{\prime}=\sum_{\vec k\eta}\left(\frac{\hbar^2}{2m}
k^2+\eta\left(\frac{\hbar^2}{m}\vec Q\vec k+\frac{\Delta}{2}\right)\right)
c^+_{\vec k\eta}c_{\vec k\eta}
\label{ham5}
\end{equation}
with $\Delta$ being the Zeeman gap.
This alteration breaks the SU(2) symmetry down to U(1) as the
Hamiltonian of course still commutes with $\Sigma$ but not with 
$T^{\pm}_{\vec Q^{\prime}}$ for any choice of $\vec Q^{\prime}$ since
\begin{eqnarray}
& & \left[{\cal H}^{\prime},T^{\pm}_{\vec Q^{\prime}}\right]\nonumber\\
& & =\sum_{\vec k}
\left(\frac{2\hbar^2}{m}\vec k\left(\vec Q-\vec Q^{\prime}\right)+\Delta\right)
c^+_{(\vec k-\vec Q)+}c_{(\vec k+\vec Q)-}\,.
\label{commT2}
\end{eqnarray}

The operators $T^{\pm}_{\vec Q}$ are defined with respect to the explicit
spinors (\ref{eigenspinor}). In terms of the usual spin density
$\vec S_{\vec q}=(1/2)\sum_{\vec k}\sum_{\mu\nu}c^+_{\vec k\mu}\vec\sigma_{\mu\nu}c_{(\vec k+\vec q) \nu}$
defined with respect to the original spin coordinates underlying the
Hamiltonian (\ref{ham1}) they can be expressed as
\begin{equation}
T^{\pm}_{\vec Q}=S^z_{\pm 2\vec Q}\pm i\frac{\vec Q}{|\vec Q|}
\cdot\vec S_{\pm 2\vec Q}\,,
\end{equation}
i.e. they describe the spin components perpendicular to the quantization axis
defined by $\Sigma=2T^3$. Defining the hermitian combinations
$T^1=(T^++T^-)/2$, $T^2=(T^+-T^-)/(2i)$, we obtain an su(2)-valued
vector $\vec T$ of observables commuting with the Hamiltonian. Thus, the
expectation value $\langle\vec T\rangle$ within any pure state is constant
in time. Regarding mixed states, a sufficient condition for a constant
expectation value is to demand that the density operator is only a function
of the Hamiltonian itself, $\rho=\rho({\cal H})$, as typical for equilibrium
situations. However, such a density
matrix is also invariant under arbitrary spin rotations generated by
$\vec T$ such that, as usual for rotationally invariant magnetic systems, 
$\langle\vec T\rangle$ vanishes. In other words, a finite expectation value
$\langle\vec T\rangle$ is the consequence of a non-equilibrium state or
the result of spontaneous symmetry breaking. The latter should of course
not be expected in a two-dimensional system.

For quantum wells with growth direction [110] the Dresselhaus term
(\ref{dressel110}) commutes with $\sigma^z$. Thus, the analog of the 
Hamiltonian (\ref{ham1}) allows for a conserved spin quantity 
if Rashba spin-orbit coupling is absent. In this case the Hamiltonian
can again, neglecting the cubic third-harmonic contribution to the
Dresselhaus coupling, be formulated as in Eq.~(\ref{ham2}) with 
\begin{equation}
\Sigma=\sigma^z\qquad,\qquad
\vec Q=\frac{m\beta}{2\hbar^2}(0,1)\,.
\label{sigmaq2}
\end{equation}
With these replacements, analogous properties as obtained above
for [001] quantum wells at $\alpha=\pm\beta$ follow. In particular, 
an SU(2) symmetry as described in Eqs.~(\ref{ham4})-(\ref{commrho})
also occurs here which is in the present case broken down to U(1) by
applying a magnetic field along the growth direction (cf. Eqs.~(\ref{ham5}),
(\ref{commT2})).

Finally, for quantum wells grown into the [111]-direction the
linear part of the Dresselhaus coupling (\ref{dressel111}) and the Rashba
term (\ref{rashba}) have the identical functional form. Here a conserved
quantity can only be realized if these two contributions exactly cancel
each other \cite{Cartoixa05a,Cartoixa05b,Vurgaftman05}. 

Moreover, in a very recent work \textcite{Kammermeier16} extended the
above considerations to more general growth directions. Specifically
it was demonstrated that a conserved spin operator of the above type
exists, for appropriately tuned Rashba coupling, {\em if and only if two
Miller indices of the growth direction agree in modulus.} Fully analogously
to the cases discussed above, this conserved spin components are extended
to an su(2) algebra of operators commuting with the Hamiltonian.

%\vspace{0.5cm}
\subsection{Spin Diffusion Equations}
\label{sdiff}

Let us concentrate again on [001] quantum wells.
As seen before, for balanced contributions to
spin-orbit coupling $\alpha=\pm\beta$ an electron spin undergoes a
perfectly controlled rotation provided the locations of injection and detection
of the electron are sufficiently defined, for instance
in terms of quantum point contacts \cite{Schliemann03a}. This, however, is a
rather special situation in experiments. In order to treat more general
scenarios it is useful to study the expectation value of the local spin
density,
\begin{equation}
\vec s(\vec r,t)=\left\langle\sum_a\frac{\hbar}{2}\vec\sigma_a
\delta\left(\vec r-\vec r_a(t)\right)\right\rangle\,,
\end{equation}
where $a$ labels the electrons, and the average $\langle\cdot\rangle$ is to be
taken over the given (in general nonequilibrium) state in the presence of
disorder potentials and/or interactions among the charge carriers. Moreover,
we also include the cubic third-harmonic correction to the Dresselhaus term
(\ref{dressel001-1}) proportional to $\beta_3$. Effective
semiclassical diffusion equations for $\vec s(\vec r,t)$ can be derived
via quantum kinetic equations rooted in the Keldysh formalism
\cite{Mishchenko04}. Working in Fourier space at small frequencies and
wave vectors, and evaluating the arising parameters within the zero-temperature
ground state, one obtains in the regime of weak spin-orbit coupling
\cite{Bernevig06,Stanescu07,Liu12,Salis14}
\begin{equation}
\left(-i\omega+Dq^2+D_{\rm so}\right)
\left(
\begin{array}{c}
\bar s^1(\vec q,\omega) \\
\bar s^2(\vec q,\omega) \\
\bar s^3(\vec q,\omega)
\end{array}
\right)=0
\label{diffeq1}
\end{equation}
with 
\begin{widetext}
\begin{equation}
D_{\rm so}=2k_f^2\tau\left(
\begin{array}{ccc}
\left((\alpha+\beta)^2+\beta_3^2\right)/ \hbar^2 & 0 
& i(\alpha+\beta)\bar q_1/m \\
0 & \left((\alpha-\beta)^2+\beta_3^2\right)/ \hbar^2 
& i(\alpha-\beta)\bar q_2/m \\
-i(\alpha+\beta)\bar q_1/m & -i(\alpha-\beta)\bar q_2/m
& 2\left(\alpha^2+\beta^2+\beta_3^2\right)/ \hbar^2
\end{array}
\right)\,.
\label{matD}
\end{equation}
\end{widetext}
In Eq.~(\ref{diffeq1}) we have used a rotated coordinate system in the
plane of the quantum well, $\bar s^{1,2}=(\pm s^x+s^y)/\sqrt{2}$,
$\bar s^3=s^z$,  and likewise for the wave vector $\vec q$, such that
the new axes are along [110] and [1$\bar 1$0]. $D=v_f^2\tau/2$
is the usual diffusion constant given in terms of the momentum relaxation time
$\tau$ and the Fermi velocity $v_f=\hbar k_f/m$ for an effective Fermi
wave vector $k_f$ (neglecting again spin splitting here).
The above result (\ref{diffeq1}) is valid in the regime of weak 
spin-orbit interaction, $(\alpha,\beta,\beta_3)k_f\tau/ \hbar\ll 1$.
Diffusion equations similar to Eqs.~(\ref{diffeq1}),(\ref{matD}) have,
for various types of spin-orbit coupling and ingredients to the many-body
physics, also been derived using different theoretical techniques
\cite{Bernevig08,Burkov04,Kalevich94,Kleinert07,Kleinert09,Luffe11,Luffe13,Raimondi06,Schwab06,Wenk10,Yang10}.

\begin{figure}
\includegraphics[width=\columnwidth]{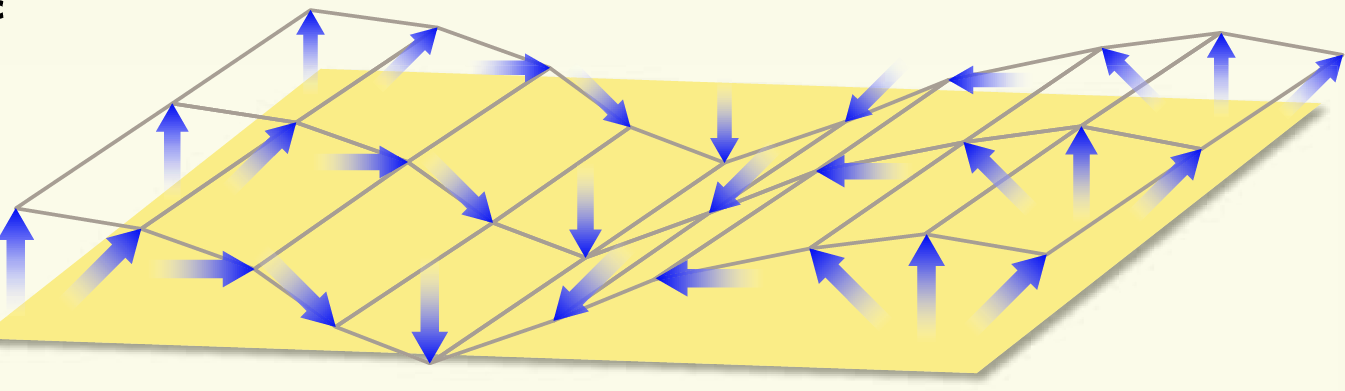}
\vspace{0.2cm}
\caption{(Color online)
Schematic of the persistent spin helix occuring in a [001] quantum well
for spin-orbit coupling tuned to $\alpha=\beta$ according to 
Eq.~(\ref{diffpss}). The shift vector $\vec Q$ defines the pitch of the helix
and points along the lateral direction. The spin density component
in the longitudinal direction vanishes, $\bar s^2=0$. 
Adapted from \textcite{Fabian09}.}
\label{Fabian}
\end{figure}
All the effects of spin-orbit interaction in the diffusion equation 
(\ref{diffeq1}) are encoded in the matrix (\ref{matD}).
As to be expected, this equation also
reflects the symmetry properties arising for balanced Rashba and Dresselhaus
coupling as analyzed in Sec.~\ref{pshbasic}: At, say, $\alpha=\beta$ the
equation for $\bar s^2$ decouples from the remaining system and reads in real
space
\begin{equation}
\left(\partial_t-D\nabla^2+1/T_1\right)
\bar s^2(\vec r,t)
\end{equation}
with $1/T_1=2k_f^2\tau(\beta_3/ \hbar)^2$ and the general solution
\begin{equation}
\bar s^2(\vec r,t)=\frac{e^{-t/T_1}}{(2\pi)^2}
\int d^2qe^{-Dq^2t}\bar s^2(\vec q,t=0)e^{i\vec q\vec r}\,.
\end{equation}
The latter equation describes the diffusion of an initial spin polarization,
accompanied by its
decay on the time scale $T_1$ which, as suggested by the notation, is aptly
referred to as a decoherence time. Without the cubic third-harmonic
correction to the Dresselhaus term, $\beta_3=0$, no decay occurs, and all the
dynamics is due to the diffusive motion of electrons with fixed spin governed
by the particle (or charge) diffusion constant $D$. 
We note that the spin density can be changed by either moving the particles or
altering their spin. Due to the latter mechanism, the spin density does,
differently from the charge density, fulfill a continuity equation
with additional source terms \cite{Erlingsson05}.  
The infinite spin relaxation time occurring at $\alpha=\pm\beta$ and
$\beta_3\propto 1/T_1=0$ was confirmed by several authors on the basis
of Monte Carlo simulations treating the orbital carrier dynamics classically
\cite{Kiselev00a,Ohno07,Ohno08,Liu10}; for an analytical approach see also
\textcite{Lyubinskiy06,Wenk11}.

The two other solutions to Eq.~(\ref{diffeq1}) are
for $\alpha=\beta$ and $\beta_3=0$ characterized by the frequencies
\begin{equation}
i\omega_{\pm}(\vec q)=Dq^2\pm8k_f^2\tau
\left((\alpha/ \hbar)^2+(\alpha\bar q_1/m)^2\right)\,.
\end{equation}
Now for $\vec q$ being twice the ``shift vector'' $\pm\vec Q$ as occurring
in Eqs.~(\ref{sigmaq}),(\ref{degrel}), i.e. $\bar q_1=\pm2|\vec Q|$,
$\bar q_2=0$, we have $\omega_-=0$ \cite{Bernevig06}. 
This static solution describes the
persistent spin helix and reads in real space
\begin{equation}
\left(
\begin{array}{c}
\bar s^1(\vec r) \\
\bar s^3(\vec r)
\end{array}
\right)
=A\left(
\begin{array}{c}
\cos\left(2\vec Q\vec r+\phi\right) \\
-\sin\left(2\vec Q\vec r+\phi\right)
\end{array}
\right)
\label{diffpss}
\end{equation} 
with two real constants $A$, $\phi$. Naturally, the angular argument
$2\vec Q\vec r$ of the spin rotation around the [1$\bar 1$0] direction 
occurring here is the same as (for $\vec r=\vec a$)
in the effective evolution operator (\ref{timeev}): The spatial dependence
of the spin density (\ref{diffpss}) precisely mimics the rotation of the
spin of an electron moving along the direction of $\vec Q$.
Fig.~\ref{Fabian} shows a sketch of the helical spin structure
described be Eq.~(\ref{diffpss}). 

\subsection{[001] Quantum Wells: Experiments and Simulations} 
\label{expsim}

We now review experimental work investigating the high-symmetry situation
$\alpha=\pm\beta$ in [001] quantum wells, along with numerical simulations
and pertinent theoretical approaches.

\subsubsection{Optical Techniques}
\label{opttech}
\begin{figure}
\includegraphics[width=\columnwidth]{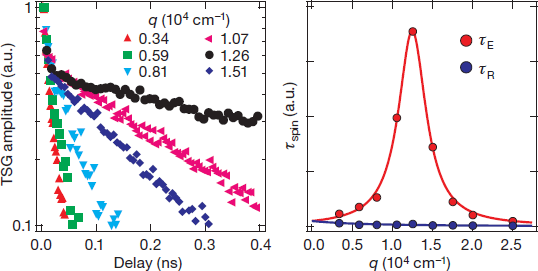}
\includegraphics[width=\columnwidth]{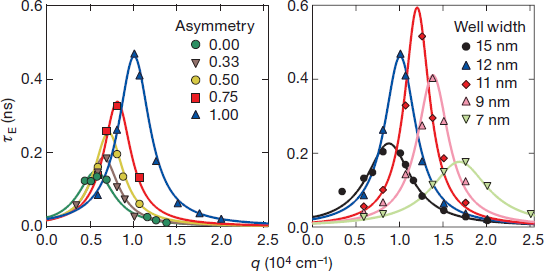}
\vspace{0.2cm}
\caption{(Color online) Upper left panel:
Decay curves of spin gratings obtained by \textcite{Koralek09}
for different wave numbers $q$ of the initially modulated spin density.
The data shows decay on two distinct time scales.
Upper right panel: Lifetimes of the spin helices with enhanced
($\tau_E$) and reduced ($\tau_R$) stability. The former one
corresponds to the spin density configuration (\ref{diffpss}) with
a maximum lifetime at $q=2Q$.
Lower panels: Lifetime of the spin helix configuration as a function
of wave number for different doping asymmetry (varying the
Rashba coupling, left) and well width (varying the Dresselhaus term, right).
Adapted from \textcite{Koralek09}}.
\label{KoralekEtAl}
\end{figure}

The stability of periodic spin structures in GaAs quantum wells close
to the regime $\alpha\approx\beta$ has been experimentally studied by
\textcite{Koralek09} using the technique of transient spin-grating
spectroscopy. Here a periodic spin pattern with defined wave vector $\vec q$
is created via optical orientation by two noncollinear laser beams, and its
time evolution is then monitored by diffraction of a time-delayed probe pulse. 
The initial spin density structure is a superposition of two helices
with the same $\vec q\parallel[110]$ but different senses of rotation, 
only one of which matches the one encoded in the static solution 
(\ref{diffpss}).
Accordingly, \textcite{Koralek09} observe that the initial spin polarization
decays, to about equal weights, on two very distinct time scales
as shown in Fig.~\ref{KoralekEtAl}:
A short-lived part where the life time shows a maximum at
$q=0$ and slowly decreases with growing wave number, and
a fraction with clearly enhanced lifetime attaining a pronounced maximum
at $q\approx 10^6{\rm m}^{-1}$. The latter should be interpreted as a
persistent spin helix (\ref{diffpss}) with $q=2Q=4m\alpha/ \hbar^2$
corresponding to a Rashba parameter of 
$\alpha\approx 3{\rm meV}\AA$. The fact that these measurements
indeed explore the regime $\alpha\approx\beta$ was further established
by varying the Rashba and Dresselhaus parameter
(see lower panels of Fig.~\ref{KoralekEtAl}): The first was achieved by
studying samples with altered relative concentration of the remote
dopants on both sides of the quantum well at fixed total density of
dopants, while in the latter case samples of different well width were
compared.

In a related experimental study \textcite{Yang12} investigated, also
using transient spin-grating spectroscopy, the drift dynamics of spin helices in
the presence of an electric field directed in the plane of a {\em symmetric}
quantum well with a vanishing Rashba but finite Dresselhaus parameter.
These spin helices have of course per se a finite lifetime.
\begin{figure}
\includegraphics[width=\columnwidth]{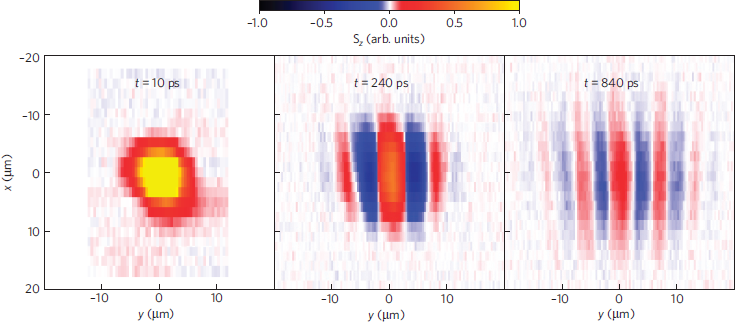}
\includegraphics[width=\columnwidth]{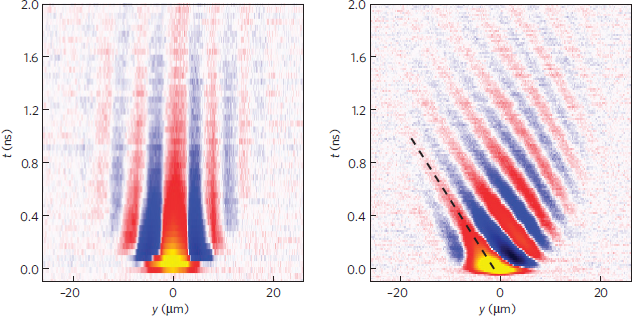}
\vspace{0.2cm}
\caption{(Color online)
Kerr rotation data obtained by \textcite{Walser12}. The detection
method is sensitive to the out-of-plane component $s^z$ of the spin density. 
Upper panels:
Time evolution of the persistent spin helix from an initial local spin
polarization along the growth direction (left) generated by a pump laser.
Here the $x$- and $y$ direction point along [1$\bar 1$0] and [110], 
respectively, and the detection method is only sensitive to the out-of-plane
component $s^z(\vec r,t)$ of the spin density shown.
Lower panels: 
Time evolution of the spin density in zero magnetic field
(left) and in an in-plane field of B=-1{\rm T} along [1$\bar 1$0] rotating
the in-plane spin components into the growth direction.
Adapted from \textcite{Walser12}.}
\label{WalserEtAl}
\end{figure}

The formation of a persistent spin helix was directly observed by
\textcite{Walser12} using spatially and temporally resolved Kerr microscopy.
The authors monitored the time evolution of a
local spin polarization along the growth (or $z$-)direction 
produced by a focused
pump laser. The upper panels of 
Fig.~\ref{WalserEtAl} show the spreading of this initial 
wave packet by diffusion: The 
$z$-component of the spin density evolves, due to the combined Rashba and
Dresselhaus spin-orbit coupling near $\alpha=+\beta$,
into an oscillatory pattern along the [110]-direction, consistent with
numerical simulations by \textcite{Liu06,Liu09}.
As to be expected,
the spin density pattern is constant in the orthogonal 
[1$\bar 1$0]-direction. Similar as in the measurements 
by \textcite{Koralek09}, the
detection technique is only sensitive to the $z$-component of the
spin density. Applying an external magnetic field
in [1$\bar 1$0]-direction coupling to the conserved spin component $\Sigma$ 
rotates the in-plane component of the helix into the growth direction and
enables its detection (lower panels of Fig.~\ref{WalserEtAl}). 
We note that, as seen in Eqs.~(\ref{ham4}),
(\ref{commT2}), introducing such an external field breaks the SU(2)
symmetry at $\alpha=\beta$. Thus, the work by \textcite{Walser12} is
an experimental demonstration that the stability of the persistent spin
helix does not depend on the full SU(2) symmetry but the
existence of the single conserved spin component $\Sigma$ suffices.

The investigations by \textcite{Walser12} were performed in a [001]-grown GaAs
quantum well with the Rashba parameter fixed by asymmetric doping.
\textcite{Ishihara14a} have conducted a similar imaging study on a
sample where the Rashba parameter was varied by a gate voltage close
to $\alpha=-\beta$. In a companion study the same authors used 
Kerr imaging to map
out the spin dynamics in quantum wires lithographically defined in the 
quantum well along the [110]- and [1$\bar 1$0]-direction
\cite{Ishihara14b}. In accordance with Eqs.~(\ref{ham2}), (\ref{sigmaq})
(again for $\alpha=-\beta$),
the spin-orbit coupling was in the former case 
(where $\vec k\perp\vec Q$ ) found to be strongly
suppressed while in the latter direction a spin helix was formed. 
Similar results on spin dephasing time scales 
for quantum wires in GaAs wells close to
$\alpha=-\beta$ were reported by \textcite{Denega10} . 

\textcite{Salis14} combined the experimental techniques of
\textcite{Walser12} with theoretical simulations to study the formation
of a spin helix, again following a local optical spin excitation, 
under imperfect conditions. Specifically, the authors considered 
a finite imbalance $|\alpha|-\beta\neq 0$, a substantial third-harmonic cubic
correction to the Dresselhaus term, and lateral confinement within the
quantum well. The experimental results obtained again for GaAs samples
are found to agree well with the theoretical modeling. 
Only shortly later the same collaboration \cite{Altmann14}
investigated, in a similar
experimental setup, the spin helix lifetime near $\alpha=\beta$
in  quantum wires etched along the direction [110]$\parallel\vec Q$
in quantum wells originating from the same wafer as used before
\cite{Walser12}. By fitting their data to a spin diffusion model
(\textcite{Salis14}, cf. Sec.~\ref{sdiff}) the authors conclude that
the observed enhanced stability of the helix is mainly due to the
geometrical confinement while the intrinsic lifetime is rather unaffected
and still determined essentially by the cubic third-harmonic contribution
to the Dresselhaus term.

\textcite{Schonhuber14} have investigated via inelastic light scattering
intrasubband spin excitations in a GaAs quantum well close to
$\alpha=\beta$ produced again from the same wafer as used by
\textcite{Walser12}. For momentum transfer along the direction of
$\vec Q\parallel$[110] a substantial spin splitting is found, while this
quantity is clearly suppressed in the opposite direction, in accordance with 
Eqs.~(\ref{ham2}), (\ref{sigmaq}) and the findings by \textcite{Ishihara14b}
(working at $\alpha=-\beta$). The spin orbit parameters extracted 
from the measurements are consistent with the results by \textcite{Walser12}.

Most recently, two studies extended the work by \textcite{Yang12} mentioned 
above on spin helix drift in quantum wells now close to $\alpha=\pm\beta$.
\textcite{Kunihashi16} investigated drift spin transport via Kerr imaging
in a four terminal geometry of ohmic contacts covered by a semi-transparent
Au gate electrode. The latter varied the Rashba coupling while voltages applied
to the contacts created drift transport of optically injected 
spin-polarized  electrons.
Wells of two widths ($L=15{\rm nm}$ and $L=25{\rm nm}$)
were studied with the wider one being close to $\alpha=-\beta$, 
and a clearly enhanced spatial coherence of the drifting spin pattern was 
observed here. The authors also demonstrated the modulation of the
electron transport path upon applying time-dependent drift voltages.
\textcite{Altmann16} used samples of the same structure as \textcite{Walser12}
to perform a Kerr imaging study concentrating on the situation where
the diffusive current of the optically injected spin density is compensated
by the drift current. Here a spin precession is found with a frequency
proportional to the drift velocity. Using an appropriate model for the
carrier distribution function (being anisotropic as a function of wave
vector) the authors explain this effect with properties of the
cubic Dresselhaus term.

\subsubsection{Transport Measurements}
\label{transport}
\begin{figure}
\includegraphics[width=\columnwidth]{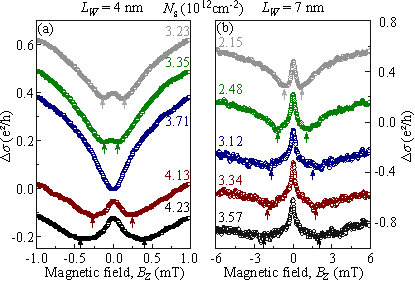}
\vspace{0.2cm}
\caption{(Color online) Magnetoconductance as measured by \textcite{Kohda12}
for different gate voltages in two quantum wells differing in width.
In the narrower well (left) a transition from weak antilocalization
to weak localization and back occurs. 
From \textcite{Kohda12}.}
\label{KohdaEtAl}
\end{figure}

\textcite{Kohda12} have investigated the quantum corrections to the
magnetoconductance in InGaAs quantum wells with spin-orbit
coupling close to $\alpha=\beta$. Spin-orbit interaction 
combined with scattering on imperfections generically
randomizes the spin leading to weak antilocalization signaled by a 
negative magnetoconductance \cite{Knap96,Wirthmann06,Schapers06,Kettemann07}. 
For Rashba and linear Dresselhaus spin-orbit interaction at $\alpha=\pm\beta$,
however, spins are left unaltered along closed trajectories which should give
rise to weak localization, i.e. a positive magnetoconductance.
\textcite{Kohda12} applied the magnetic field perpendicular to the
quantum well and used a gate voltage across it to vary the Rashba parameter
$\alpha$.
As shown in Fig.~\ref{KohdaEtAl},
in a comparatively narrow well of width $L=4{\rm nm}$ the authors found
a transition from weak antilocalization to weak localization and back when
driving the gate voltage through an appropriate critical value. In a 
sample with a larger well width of $L=7{\rm nm}$ and therefore smaller
Dresselhaus parameter $\beta$ no such behavior was observed, i.e. in the latter
system $\beta$ and the range of $\alpha$ seem to be too different to match
each other.
The experimental results are corroborated by numerical simulations which
conclude that the weak localization signal persists
even if the cubic third-harmonic term in Eq.~(\ref{dressel001-1}) 
is included, but its location in parameter space is shifted away from
$\alpha=\beta$.

The fact that the $L=4{\rm nm}$ sample 
actually has spin-orbit coupling parameters close to $\alpha=\beta$ 
was also established by \textcite{Kohda12} in an independent experiment 
using the spin-galvanic effect. This phenomenon amounts in an electric current
to response to an in-plane spin polarization, and its directional dependence
is highly sensitive to the ratio $\alpha/ \beta$ 
\cite{Ganichev04,Ganichev14,Trushin07b},
which was indeed found to be close to unity.

\textcite{Dettwiler14} extended the investigations by \textcite{Kohda12}
using GaAs quantum wells varying in width from $L=8{\rm nm}$ to
$L=13{\rm nm}$. Employing a combination of top and back gates the Rashba 
parameter $\alpha$ and the carrier density $n$ 
could be tuned independently. The point
$\alpha=\beta$ was again determined by monitoring the transition from
weak antilocalization to weak localization and back. To obtain a 
consistent data analysis it was necessary to take into account the
cubic correction (being proportional to $n$) to the Dresselhaus parameter
$\beta$. As a result, \textcite{Dettwiler14} demonstrated control over
spin-orbit coupling parameters and carrier density while preserving the
condition $\alpha=\beta$. At quite high densities such as 
$n=9\cdot 10^{11}{\rm cm}^{-2}$ no transition between weak antilocalization and
localization was found which should be ascribed to the third-harmonic
correction to the Dresselhaus term that also increases with density.

Magnetoconductance studies in quantum wires in the directions [100],
[110], [1$\bar 1$0] of an [001] InGaAs quantum well were performed
by \textcite{Sasaki14} building upon theoretical work by \textcite{Scheid08}. 
To reduce fluctuation effects the authors used arrays of wires
which were arranged in the same sample thus enabling simultaneous
measurements. The Rashba coupling was varied by a top gate, and the
magnetic field lay in the plane of the well leading in combination with 
the spin-orbit coupling to a strong anisotropy of the magnetoconductance
which additionally depends on the direction of the wire.
In a one-dimensional quantum wire spin randomization due to momentum
scattering (Dyakonov-Perel mechanism) is quenched since  
effective wave vector-dependent field
provided by the spin-orbit interaction is unidirectional.
This phenomenon will be  discussed in more detail in section
\ref{lateral}.
An in-plane magnetic field noncollinear with the spin-orbit field
changes this situation and leads spin randomization favoring weak
antilocalization. Thus,
in accordance with numerical simulation done by the authors, 
the weak localization signal in the conductance is maximal if the
applied field is collinear with the field
provided by the spin-orbit coupling. With the direction of the wave vector
defined by the quantum wire, the latter observation provides a 
means to determine the ratio $\alpha/ \beta$. In particular, for
$\alpha/ \beta=1$ no anisotropy of the magnetoconductance is observed
for transport in the [1$\bar 1$0]-direction since here we have 
$\vec k\perp\vec Q$ (cf. Eqs.~(\ref{ham2}), (\ref{sigmaq})). 
If the applied magnetic field is substantially noncollinear with the
spin-orbit field it randomizes the spin and suppresses weak localization.
Assuming that this process is most efficient if both fields are of the
same magnitude (as indicated by the numerics), \textcite{Sasaki14} 
also give reasonable separate estimates for $\alpha$ and $\beta$.
For a further proposal to determine the relative strength of the
Rashba and Dresselhaus coupling utilizing the high-symmetry point
$|\alpha/ \beta|=1$ see \textcite{Li10}.

Most recently, \textcite{Yoshizumi16} have demonstrated gate-controlled
switching between $\alpha=\pm\beta$ in an InAlAs quantum well.
The occurrence of each persistent spin helix (differing by sense of rotation)
was again detected by similar magnetoconductance measurements as above.

\subsubsection{Stability of the Spin Helix: Limiting Factors}
\label{limitations}

Several theoretical studies have identified the cubic Dresselhaus term
as the main decay mechanism of the persistent spin helix
\cite{Lusakowski03,Cheng06,Cheng07,Kettemann07,Shen09,Wenk11,Luffe11,Liu12,Kurosawa15},
in accordance with experiments already mentioned \cite{Koralek09,Salis14}.

Specifically, the spin-grating experiments by \textcite{Koralek09} found the
spin lifetime $\tau_h$ of the symmetry-protected spin helix
to be of order a few hundred picoseconds, depending
significantly on temperature. The ratio of $\tau_h$ and the time scale
of ordinary (spin) diffusion can be expressed as $\eta:=4DQ^2\tau_h$.
This quantity is approximately constant, $\eta\approx 100$, below $50{\rm K}$
while it decreases for higher temperature with a power law showing an
exponent slightly larger than $2$. The subsequent theoretical analysis
by \textcite{Liu12} concluded that this temperature dependence 
cannot be quantitatively described by a low-order treatment
of the spin-orbit interaction which is essentially restricted to the
Dyakonov-Perel regime and leads to the diffusion equations (\ref{diffeq1}),
(\ref{matD}) (given here at zero temperature). This finding is in 
qualitative agreement with the experimental study by \textcite{Studer09}
on InGaAs quantum wells using time-resolved Faraday rotation.
Instead the Elliot-Yafet
relaxation mechanism should also be taken into account which yields expressions
somewhat more involved than Eqs.~(\ref{diffeq1}), (\ref{matD}).

Theoretical investigations by \textcite{Luffe11,Luffe13} led to the
prediction that the spin helix life time can be enhanced by
Coulomb repulsion (treated there within Hartree-Fock approximation).
A study of Rashba and Dresselhaus coupling and its interplay with Coulomb
interaction described by the GW approximation was presented by 
\textcite{Nechaev10}.

A further possible source of decoherence of the spin helix are spatial 
inhomogeneities of the effective Rashba coupling 
\cite{Liu06,Glazov10b,Bindel16}.

\subsection{Many-Body Signatures of the Persistent Spin Helix}
\label{manybody}

We now summarize the role of persistent spin textures in connection with the
many-body physics of interacting systems. If not mentioned otherwise,
we consider electrons in [001] quantum wells 
subjected to Coulomb repulsion and spin-orbit coupling of the
Rashba and the linear Dresselhaus type.

\textcite{Badalyan09,Badalyan10a} evaluated the dielectric function of
the two-dimensional electron gas within Random Phase Approximation 
(Lindhard formula). For $\alpha=\pm\beta$ one obtains the dielectric response of
system without spin-orbit coupling, while for general parameters 
a beating of the static Friedel oscillations is observed. In a subsequent
work the charge density relaxation propagator, i.e. the slope of the
imaginary part of the polarization function, and its analyticity properties
was studied \cite{Badalyan13}.

The optical conductivity was calculated by \textcite{Maytorena06} and
compared with the frequency dependence of the spin Hall conductivity
(which vanishes in the static limit \cite{Schliemann06}).
The authors predict a rich phenomenology arising from the interplay of
the two spin-orbit coupling terms. In a subsequent work the analysis
was extended to the optical (i.e. spatially homogeneous) spin 
susceptibility \cite{Lopez-Bastidas07}.
The optical conductivity for quantum wells with Rashba and Dresselhaus coupling
was reconsidered by \textcite{Li13}. As a signature of the persistent
spin helix, all interband transitions vanish at $\alpha=\pm\beta$. If the
cubic Dresselhaus contribution is taken into account, these transitions
are rendered finite but still suppressed.

\textcite{Capps15} studied the finite-temperature equilibrium state
of an interacting electron gas at $\alpha=\pm\beta$ within Hartree-Fock
approximation and concluded the absence of any helical spin structures;
a finding consistent with the fact that, as discussed in section 
\ref{pshbasic}, finite expectation values
of the operators (\ref{T3}), (\ref{Tplus}) occur only in nonequilibrium states
or as the result of spontaneous symmetry breaking. Most recently, the
authors have extended their analysis to the spin Seebeck effect
\cite{Capps16}.

The RKKY interaction between magnetic
moments in the presence of spin-independent disorder was investigated by
\textcite{Chesi10}. Here the 
disorder-averaged susceptibility shows a
twisted exchange interaction decaying exponentially with distance.
\textcite{Iglesias10} have investigated the dynamical spin-polarization, i.e
the linear response of the spin-magnetization to a homogeneous
in-plane electric field. The authors consider Rashba and Dresselhaus spin-orbit
coupling in quantum wells with growth directions [001], [110], and [111].

When the electrons are confined to a quantum wire (and their interaction is
neglected) the spin-orbit coupling in general leads to
anticrossings of the single-particle subband dispersions except for the
case $\alpha=\pm\beta$ where, due to the additional conserved quantity,
crossings occur \cite{Schliemann03a}. Using a Luttinger liquid description,
\textcite{Meng14} studied the renormalization of such (anti-)crossings
in the presence of Coulomb repulsion. This effect especially significant
near the high-symmetry point $\alpha=\pm\beta$ where the anticrossing
gap vanishes with an interaction-dependent power law in the spin-orbit
parameters.

\subsection{Other Growth Directions and Geometries}
\label{ogrowgeo}

We now review, among other items, experimental studies dedicated to persistent
spin structures in quantum wells of the other high-symmetry growth
directions [110] and [111]. A very recent prediction of analogous phenomena
in systems of more general growth direction has been already mentioned 
\cite{Kammermeier16}. For a summary of experimental work on spin-orbit
coupling in such systems (not specifically addressing spin helices)
we refer to \textcite{Ganichev14}.

\subsubsection{[110] Quantum Wells}

As seen in Sec.~\ref{pshbasic}, for quantum wells grown in the [110]-direction
a conserved spin component along with an SU(2) symmetry involving an
appropriate wave vector transfer occurs in the absence of Rashba coupling. 
Relying on optical techniques several groups reported on clearly enhanced
spin dephasing times compared to those observed
in quantum wells of other growth directions
\cite{Ohno99,Schreiber07a,Schreiber07b,Cuoto07,Belkov08,Muller08,Volkl11,Volkl14}. 
Moreover, spin dephasing is
found to be strongly anisotropic depending on whether the spin polarization 
lies in the plane of the quantum well or along the growth direction where 
the longest lifetimes occur \cite{Dohrmann04,Griesbeck12}.
These observations are of course in agreement with the structure of the
Dresselhaus spin-orbit coupling, and the remaining spin decay can be attributed
to residual Rashba coupling \cite{Tarasenko09,Glazov10a,Poshakinskiy13}
and/or hole-mediated processes \cite{Volkl11}.

\begin{figure}
\includegraphics[width=\columnwidth]{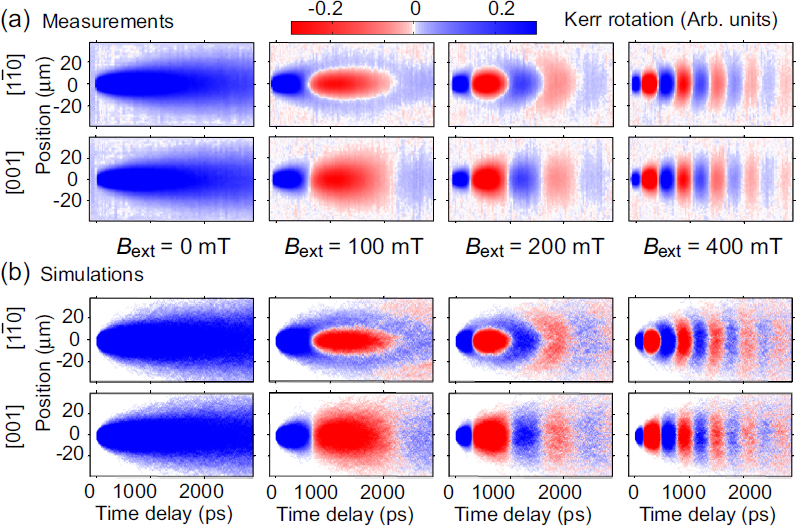}
\vspace{0.2cm}
\caption{(Color online)
Upper panels (a): Time- and spatially resolved Kerr rotation data
by \textcite{Chen14}. The dynamics of an initial spin polarization
along the [110] growth direction is followed along the [1$\bar 1$0] and
[001] direction. To generate nontrivial dynamics a magnetic field
$B_{\rm ext}$ of various strength is applied along [1$\bar 1$0].
Lower panels (b): Corresponding Monte Carlo simulation results.
From \textcite{Chen14}.}
\label{ChenEtAl} 
\end{figure}
Experiments directed explicitly towards helical spin structures were performed 
by \textcite{Chen14} who studied [110]-grown GaAs quantum wells using
time-resolved Kerr microscopy. To generate a finite net spin-orbit field
averaged over the the Fermi contour, the Fermi disk was shifted from its 
equilibrium position by applying a DC current of up to $200\mu{\rm A}$.
The direction of the current defines the direction of the effective wave 
vector to be inserted in the Dresselhaus Hamiltonian (\ref{dressel110}).
Additionally a magnetic field of order a few hundred mT was applied.
By comparing data obtained for different directions of the magnetic field, the
authors were able to extract the energy contribution due to spin-orbit
interaction. For a current along the the [1$\bar 1$0]- (or y-)direction
this quantity is proportional to the current strength, while for the
orthogonal [001]- (or x-)direction it is more or less constant, in accordance
with the form of the Dresselhaus term (\ref{dressel110}).

Similar to the studies by \textcite{Walser12} on [001] quantum wells,
\textcite{Chen14} also mapped out the formation of a helical spin structure
following a local injection of spin density polarized along the growth 
direction. As this direction coincides with the direction of the spin-orbit 
field, an additional small magnetic field was necessary to generate 
nontrivial dynamics. Fig.~\ref{ChenEtAl} shows the time-resolved data which 
is well reproduced by Monte Carlo simulations.

\subsubsection{[111] Quantum Wells}

According to Eqs.~(\ref{dressel111}) and (\ref{rashba}), the linear 
Dresselhaus coupling in quantum wells grown in the [111]-direction
can exactly cancel the Rashba term for $\alpha=2\beta/\sqrt{3}$. Thus,
spin-orbit interaction is present only in higher
corrections, the leading one being third-harmonic contribution
in Eq.~(\ref{dressel111}). This situation in GaAs quantum wells was
investigated by \textcite{Balocchi11} using time-resolved
photoluminescence spectroscopy. For an appropriate Rashba coupling 
tuned by a gate voltage, the authors observed clearly enhanced
spin lifetimes exceeding $30{\rm ns}$ for all spin directions. 
Spin polarizations perpendicular to the growth direction were
generated by a transverse magnetic field of order a few hundred mT.
In a subsequent work, \textcite{Ye12} found, for structures as used
by \textcite{Balocchi11}, the sign of the gate voltage to depend on
whether the underlying GaAs [111] substrate is terminated by
a [111]A (Ga-rich) or [111]B (As-rich) surface.

Independent confirmation for the above findings was provided by
\textcite{Biermann12} and \textcite{Hernandez-Minguez12} who performed 
photoluminescence measurements on GaAs [111] quantum wells of somewhat larger 
width; for a summary see also \textcite{Hernandez-Minguez14}.
\textcite{Wang13} recorded both the spin lifetime $\tau_s$ and the
momentum relaxation time $\tau$ in GaAs [111] quantum wells
and deduced an enhanced spin diffusion length $l_s=\sqrt{D\tau_s}$,
$D=v^2_f\tau/2$, at an appropriate gate voltage. Moreover, 
\textcite{Balocchi13} combined optical experiments and theoretical simulations
to investigate the influence of the cubic third-harmonic 
contribution to the Dresselhaus coupling close to the cancellation
of the linear part with the Rashba term. The authors conclude
that effective control over spin relaxation even at room temperature
should be possible in sufficiently narrow [111] wells where the
linear Dresselhaus term dominates.

\subsubsection{Curved Systems}

\begin{figure}
\includegraphics[width=\columnwidth]{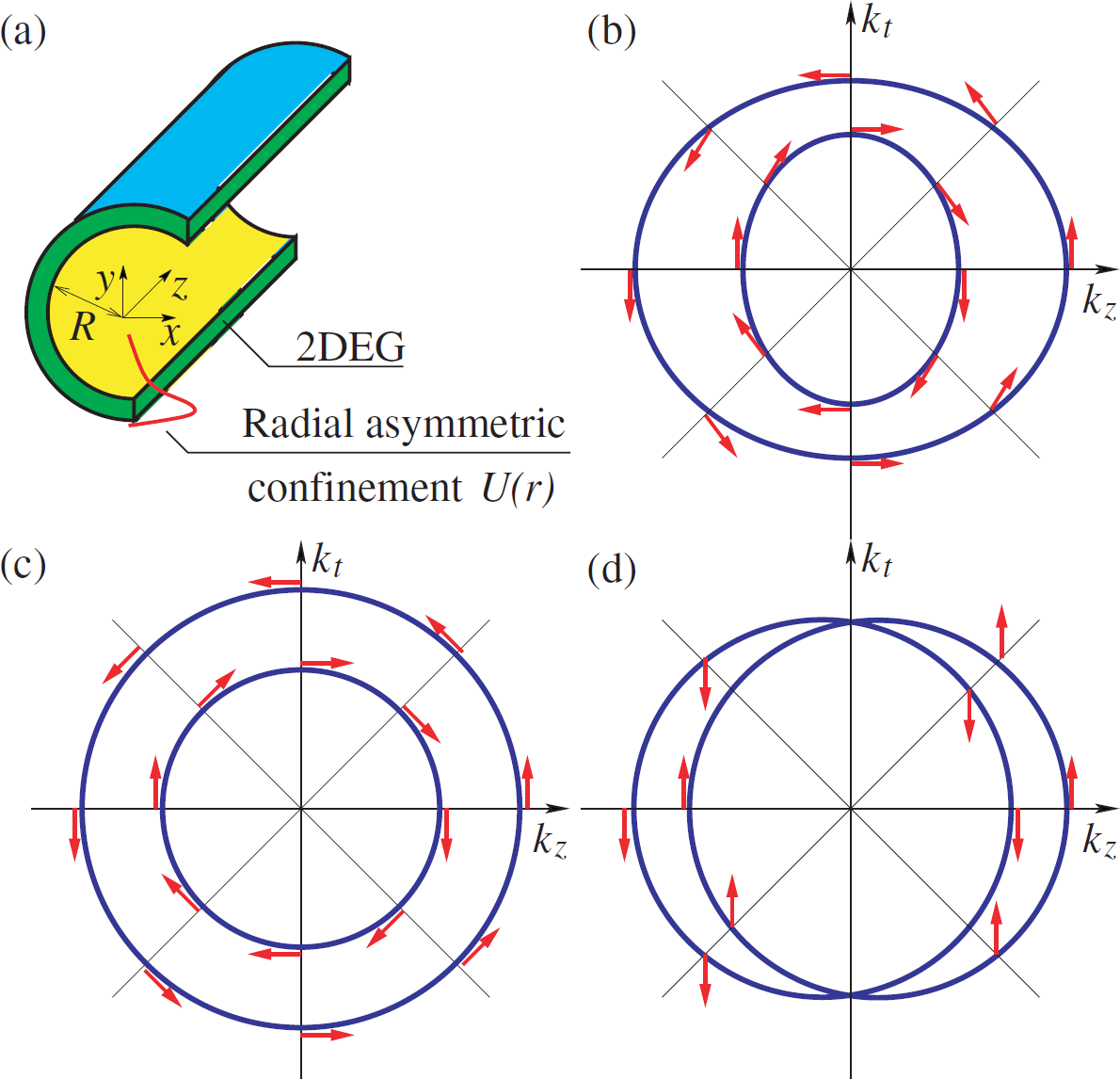}
\vspace{0.2cm}
\caption{(Color online)
(a): Sketch of a curved two-dimensional electron system
with Rashba spin-orbit coupling induecd by  asymmetric radial confinement.
(b): General case: Spin-orbit interaction leads to anisotropic Fermi
contours ($k_t=q_{\varphi}$, spin direction given by arrows).
(c): Isotropic Fermi contours of a {\em flat} system with Rashba coupling
(cf. Eq.~(\ref{rashba})).
(d) Fermi contours of the curved system with Rashba coupling tuned
according to Eq.~(\ref{radcond}): Two circles displaced by a shift vector.
From \textcite{Trushin07a}.}
\label{TrushinSchliemann} 
\end{figure}

Another situation where, for appropriately tuned spin-orbit interaction,
nontrivial conserved spin quantities occur is realized by
evenly curved cylindrical two-dimensional electron systems. The geometry
of such samples is sketched in Fig.~\ref{TrushinSchliemann}(a);
for the practical fabrication of such structures see, e.g.,
\textcite{Schmidt01,Mendach04,Mendach06}.   

Including Rashba spin-orbit coupling, the Hamiltonian can be formulated as
\cite{Trushin07a}
\begin{equation}
{\cal H}=\frac{\hbar^2k_z^2}{2m}+\frac{\hbar^2q_{\varphi}^2}{2mR^2}
+\alpha\left(k_z\sigma^{\varphi}-\frac{q_{\varphi}}{R}\sigma^z\right)
\label{curvham}
\end{equation}
where $k_z$ is the wave vector component along the ($z$-)axis of the cylinder of
radius $R$, and $q_{\varphi}=-i\partial/\partial\varphi$ generates 
real-space rotations
around the axis. $\sigma^{\varphi}=-\sigma^x\sin\varphi+\sigma^y\cos\varphi$
is the projection of the Pauli matrices on the azimuthal direction such that
$[\sigma^{\varphi},\sigma^z]/(2i)=\sigma^x\cos\varphi+\sigma^y\sin\varphi
=:\sigma^r$. For general Rashba parameter $\alpha$ the above Hamiltonian
leads to anisotropic dispersions shown in Fig.~\ref{TrushinSchliemann}(b),
differently from the case of a {\em flat} system discribed
by the Hamiltonian (\ref{rashba}) and 
depicted in Fig.~\ref{TrushinSchliemann}(c).

One easily finds the commutator
\begin{equation}
\left[{\cal H},\sigma^{\varphi}\right]=\left(\frac{\hbar^2}{2mR^2}
+\frac{\alpha}{R}\right)\left(q_{\varphi}\sigma^r+\sigma^rq_{\varphi}\right)
\end{equation}
which vanishes if the Rashba parameter fulfills
\begin{equation}
\alpha=-\frac{\hbar^2}{2mR}\,,
\label{radcond}
\end{equation}
a result that remains obviously valid if arbitrary spin-independent potentials
or interactions are added to the Hamiltonian (\ref{curvham}). Moreover, 
in full analogy to flat quantum wells
with appropriately tuned Rashba and Dresselhaus parameter, the conservation
of $\Sigma=\sigma^{\varphi}$ leads to circular dispersion relations
displaced by a shift vector (cf. Fig.~\ref{TrushinSchliemann}(d)).
Thus, analogous to Eqs.~(\ref{T3}), (\ref{Tplus}), we have 
a complete su(2) algebra of operators commuting with the Hamiltonian
\cite{Trushin07a}. Finally,
the corresponding persistent spin structure can also be described via
appropriate spin diffusion equations \cite{Kleinert09}.

Independently of the condition (\ref{radcond}) the Hamiltonian (\ref{curvham})
always commutes with the total angular momentum $j=q_{\varphi}+\sigma^z/2$,
and electrons in superpositions the same $j$
but opposite spin orientation show interesting periodic spin patterns
along the cylindrical axis \cite{Bringer11}.

As a somewhat related geometry, \textcite{Nowak09} studied 
circular quantum rings embedded in [001] quantum wells with Rashba and
Dresselhaus spin-orbit coupling. Here the latter leads, except for
the high-symmetry case $\alpha=\pm\beta$, to elliptical deformations
of the confined electron density.

\subsubsection{Lateral Confinement, Magnetic Fields, and Finite Well Width}
\label{lateral}

\textcite{Duckheim09} performed a theoretical study of
the dynamical spin Hall effect 
\cite{Duckheim07} 
in a two-dimensional electron gas confined to a channel
of finite width. Specifically, the spin accumulation at the channel boundary
in response to a AC electric field along the channel direction was investigated.
This effect is found to typically decay on the length scale set by the
spin-orbit coupling. However, considering additionally
a DC in-plane magnetic field at balanced spin-orbit coupling $\alpha=\pm\beta$
the authors were able to identify conditions under which such
spatially oscillating spin profiles can extend over the
entire channel, thus forming a driven spin helix.

\textcite{Badalyan10b} studied the interplay between Rashba and
Dresselhaus spin-orbit coupling at $\alpha=\pm\beta$ with a magnetic
field in the growth direction of the quantum well, and a hard-wall boundary
oriented either in the direction of $\vec Q$ (cf. Eq.~(\ref{sigmaq})) or
perpendicular to it. In the former case a spin helix along the boundary arises.
The key observation here is that the transformation (\ref{ansatz})
still yields a spin-independent Schr\"odinger equation (\ref{schrodingereq})
even if an arbitrary vector potential $\vec A(\vec r)$ is coupled to the
momentum $\vec p=-i\hbar\nabla\mapsto \vec p+e\vec A$, leading to a magnetic
field $\vec B(\vec r)=\nabla\times\vec A(\vec r)$ whose direct
Zeeman coupling to the spin is neglected (as done by the above authors) such 
that $\Sigma$ remains a conserved quantity. 

Introducing further lateral confinement in a given quantum well constraints the 
electron motion to take place mainly in the direction along the boundary.
An extreme case of such a situation is a (quasi-)one-dimensional quantum wire
with only the lowest subband being occupied. However, even in broader channels
the longitudinal component of the wave vector clearly dominates the dynamics. 
This constraint fixes the projection of the spin operator acting predominantly
on the electron via spin-orbit interaction. As a result, the Dyakonov-Perel
spin relaxation mechanism can be expected to be strongly suppressed by such 
lateral confinement. This effect is rather independent of particular tuning
of spin-orbit coupling parameters, but as it clearly leads to enhanced
spin lifetimes we shall also summarize the pertaining developments here;
for further discussions see also \textcite{Holleitner11}.

Following several quantitative theoretical predictions
\cite{Bournel98,Kiselev00b,Malshukov00,Pareek02}, this suppression of
spin relaxation in narrow channels was experimentally verified by
\textcite{Holleitner06,Holleitner07} combining time-resolved Faraday rotation
measurements with evaluations of Shubnikov-de Haas oscillations.
Their work was qualitatively confirmed by \textcite{Kwon07} also relying
on the Shubnikov-de Haas effect. At about the same time, the
crossover from two- to one-dimensional spin relaxation behavior was also
found by \textcite{Wirthmann06,Schapers06} via weak antilocalization 
studies.

Regarding the influence of a magnetic field in two-dimensional bulk systems
at $\alpha=\pm\beta$, \textcite{Wilde13} predicted the disappearance of 
additional beatings in de Haas-van Alphen oscillations, an effect
similar to the suppression
of {\em zitterbewegung} \cite{Schliemann06b,Nita12,Biswas12}. 

\textcite{Kunihashi09} performed magnetoconductance measurements
in narrow InGaAs wires with Rashba and Dresselhaus spin-orbit coupling 
tuned close to $\alpha=\pm\beta$. Here essentially weak localization
is found since the spin relaxation length systematically exceeds 
the inelastic scattering length due to the combined effect
of spatial dimension and the conserved spin quantity. In a subsequent
study the authors investigated a semiclassical model for
spin relaxation in systems of the above type with a focus
on $\alpha=\pm\beta$, also proposing a method to quantitatively
estimate the spin-orbit parameters \cite{Kunihashi12}.
Most recent experimental work includes investigation by
\textcite{Altmann14} on quantum wires close to $\alpha=\beta$
(cf. section \ref{opttech}) and a Kerr rotation study of channels
with dominating Dresselhaus coupling \cite{Altmann15}.

Also working close to the one-dimensional limit,
\textcite{Krstajic10} performed a theoretical study of the conductance of
quantum wires in [001] wells with spatially varying Rashba and Dresselhaus
coupling around $\alpha=\beta$ taking into account subband mixing.
Further theoretical investigations considered spin-helical structures in
quantum wires of finite length in the presence of Rashba coupling
\cite{Slipko11a,Slipko11b,Slipko13}.

\textcite{Fu15a} studied theoretically spin-orbit interaction in [001] quantum
wells broad enough such that, for typical densities, the two lowest
subbands $i=1,2$ are occupied in the ground state \cite{Bernardes07}.
Treating interaction effects within the Hartree approximation and
solving the resulting coupled Schr\"odinger-Poisson system, the authors
obtained separate coefficients $\alpha_1,\alpha_2$ and
$\beta_1,\beta_2$ for the Rashba and Dresselhaus coupling for each subband.
Points of enhanced symmetry such as $\alpha_1=\beta_1$ are discussed in more
detail. Working upon these findings \textcite{Fu15b} concluded 
that the coupling parameters can be tuned to be of equal modulus in each 
subband but with a different relative sign: $\alpha_1=\beta_1$, 
$\alpha_2=-\beta_2$. This situation gives rise to a superposition of two 
persistent spin helices with orthogonal wave vectors leading to a 
{\em persistent skyrmion lattice}.
Moreover, \textcite{Wang15} discussed the possibility of a persistent
spin helix in coupled double and triple GaAs quantum wells.

\textcite{Nakhmedov12} considered electrons subjected to Rashba and
Dresselhaus coupling near $\alpha=\beta$ and an in-plane magnetic
field in quantum wells of finite thickness where the specific
form of the transverse potential was taken into account. Moreover,
\textcite{Nazmitdinov09} investigated quantum wires with
Rashba and Dresselhaus coupling of arbitrary strength where
an in-plane magnetic field gives rise
to a conserved spin operator for electrons with appropriate longitudinal
momentum.

\subsection{Spin Field-Effect Transistors and Related Concepts}
\label{appl}

We now summarize recent developments regarding spin field-effect
transistors with a relation to persistent spin structures. For a more
comprehensive review on spin-transistor devices see \textcite{Sugahara10}.
We will mainly concentrate on [001] quantum wells.

In the classic proposal of a spin field-effect transistor due to
\textcite{Datta90} already sketched in Fig.~\ref{spinFET},
an electron is emitted from a spin-polarized electrode
into a semiconductor region where its spin is rotated via
electrically tunable Rashba coupling. Depending on the rotation angle
and the spin-polarization of the detecting electrode, the electron passes
through the device with a high or low probability, defining the
``on''- and ``off''-state of the transistor. Problems of this concept include
the spin injection into the active semiconductor region \cite{Schmidt00}
and the randomization of spins due to scattering on imperfections
as already discussed in section \ref{pshbasic} \cite{Dyakonov72}.
Indeed these obstacles limit so far the signal efficiency in
practical implementations to a rather low level \cite{Koo09}.

A means to tackle the issue of spin decay 
in [001] grown structures is to balance the 
Dresselhaus and Rashba coupling in one of the operational states of the
transistor \cite{Schliemann03a,Cartoixa03}. For example, the on-state can be
defined by $\alpha=\pm\beta$ such that the electron spin is symmetry-protected
whereas in the off-state $|\alpha|\neq\beta$ spin randomization sets in.
To fully exploit the spin conservation in the on-state, it is useful to 
precisely define the locations of spin injection and detection
by means of quantum point contacts \cite{Schliemann03a,Chuang15}. 
\textcite{Kunihashi12b} discussed a device operating between the 
states $\alpha=+\beta$ and $\alpha=-\beta$ \cite{Schliemann03a};
for a recent experimental realization of such a scenario see
\textcite{Yoshizumi16}.
A variant of the above concepts for [110] structures
was put forward by \textcite{Hall03} 
where in one of the device states additional Rashba coupling leads to
spin decoherence. Further theoretical proposals involving spatially
inhomogeneous Rashba coupling include work by \textcite{Liu12b} and
\textcite{Alomar15}.

In a proposal by \textcite{Betthausen12} electron spins are modulated
adiabatically in the on-state of the device (and therefore protected against
decay), while in the off-state diabatic Landau-Zener transitions
induced by a spatially rotating magnetic field set in leading to spin 
decoherence.
The feasibility of this device is demonstrated experimentally using a (Cd,Mn)Te 
diluted semiconductor quantum well \cite{Betthausen12}; for related
theoretical work see also \textcite{Saarikoski12,Wojcik15}.
 
\textcite{Wunderlich10} combined the spin field-transistor concept
with the spin Hall effect in an $n$-doped GaAs quantum well.
Spin are injected optically, and the conductance of the device
is switched via a top gate. The spin polarization of the resulting current
is detected via a transversal voltage resulting from the (inverse) spin
Hall effect.

\section{$p$-Doped Structures, Topological Insulators, 
and other Systems}
\label{omat}

As already sketched in Fig.~\ref{bandstructure}, the structure of the $p$-type
valence band of III-V zinc-blende semiconductors is much richer
compared to the $s$-type conduction band 
\cite{Winkler03,Fabian07,Wu10,Yu10,Korn10}.
A realistic band structure model consists here of the classic Luttinger 
Hamiltonian parametrizing the different masses for heavy and light
holes \cite{Luttinger56} and additional spin-orbit coupling contribution
arising from structure- and bulk-inversion asymmetry \cite{Winkler03}.
The latter terms are analogous to the Rashba and Dresselhaus couplings
for conduction band electrons. In particular, as pointed out only quite 
recently, in two-dimensional quantum wells pronounced Dresselhaus 
contributions being linear in the in-plane momentum
can arise from the heavy and light hole mixing induced
by boundary conditions at the nanostructure interfaces \cite{Luo10,Durnev14}.
Taking into account these findings, \textcite{Wenk16} identified
conditions for strained $p$-doped [001] wells under which conserved spin 
quantities occur {\em for holes being close to the Fermi contour}. 
The latter is for realistic parameters well approximated by a circle.
The circumstance that spin conservation applies here only to charge carriers
with wave numbers close to $k_f$ is similar to the result by
\textcite{Nazmitdinov09} who found conserved spin operators 
in $n$-doped quantum wires arising for appropriate wave vectors only.
\textcite{Wenk16} considered strained quantum wells with ground states
being both of heavy and light hole type; more special results apply
to unstrained systems. Similar conserved spin quantities in strained
quantum wells were also found by \textcite{Sacksteder14,Dollinger14}
using band structure models where the linear Dresselhaus term plays a less 
dominant role. The spin conservation here is again restricted to a vicinity
of the Fermi contour, but the necessary conditions on band structure parameters
are difficult to meet in realistic materials.

\textcite{Absor15} performed {\em ab initio} calculations for Wurtzite ZnO 
surfaces of appropriate orientation and predicted an effective spin
splitting similar as for Dresselhaus coupling $n$-doped [110] quantum wells
of zinc-blende materials (cf. Eq.~(\ref{dressel110})), which should
analogously give rise to a persistent spin helix.  

The possibility to realize persistent spin textures in monolayers of
(group-III) metal-monochalcogenides was discussed by \textcite{Li15}.

\textcite{Sacksteder12} investigated spin conduction in surface states
of three-dimensional topological insulators with anisotropic dispersion.
The authors predict coherent spin transport if (i) the effective
Hamiltonian is tuned to conserve an appropriate spin component
(such that its Dirac cone is infinitely stretched), or (ii) the
Fermi energy is aligned with a local extremum of the anisotropic
two-dimensional dispersion.

\textcite{Liu14} studied the proximity effect in layered structures
of triplet superconductors and [001] semiconductor quantum wells with
Rashba and Dresselhaus spin-orbit coupling. The authors conclude
that the vector of triplet pair expectation values should form a long-ranged 
helix in the semiconductor material for $\alpha=\pm\beta$. Related
Josephson effects and possible experimental setups are also discussed.

Yet another fascinating perspective is the possibility to realize
persistent helical spin structures in systems of ultracold fermionic atoms
as discussed by \textcite{Tokatly13}.

\section{Conclusions and Outlook} 
\label{concl}

We have reviewed the gamut of developments related to the persistent
spin helix that have emerged from the theoretical predictions by
\textcite{Schliemann03a} and \textcite{Bernevig06}. The topic is still
a hot one as one can see from the list of references, a substantial amount of
which is not older than two years. In particular, new experimental
studies continue to appear.

Our discussion includes $n$-doped
III-V zinc-blende quantum wells of growth directions
[001], [110], and [111]. The first one has received most attention so far,
regarding experiments as well as theoretical work. Right from the
beginning the investigations about conserved spin operators and suppressed
decoherence were closely tied to proposals for improving the device concept
of spin-field-effect-transistors. Among the most promising experimental
developments in this regard are the very recent studies on drift transport
of spin helices as discussed in section \ref{opttech}. It is needless to say
that these achievements have the chance to contribute to spin-based information
processing in semiconductor structures.

Also systems with growth directions [110] and [111] have been subject to
thorough experimental studies, while the predicted many-body effects
of persistent spin textures (mainly in [001] quantum wells) are still
awaiting their experimental investigation. 

A most recent theoretical result was obtained by \textcite{Kammermeier16}
stating that a persistent spin helix is achievable in quantum wells
of more general growth direction if and only if two of its Miller indices
have the same modulus. Specifically, the resulting SU(2) symmetry is
characterized by a conserved spin component $\Sigma=\vec e\cdot\vec\sigma$
and a shift vector $\vec Q$ which is always perpendicular to the
direction of the spin-orbit field, $\vec e\perp\vec Q$. The latter
feature is strongly reminiscent to the spin-momentum locking found
in the edge modes of topological insulators 
\cite{Hasan10,Qi11} but occurs here in the bulk of the system.
The above findings have the potential to stimulate
a flurry of further fascinating experiments. 

Other very recent new developments
include the study of two-dimensional $p$-doped systems which are
well known for their clearly richer band structure.

\acknowledgments

I thank E. Bernardes, G. Burkard, T. Dollinger, J.~C. Egues, S.~I. Erlingsson, 
S.~D. Ganichev, M.~S. Garelli, M. Kammermeier, M. Lee, D. Loss, K. Richter, 
D. Saraga, A. Scholz, M. Trushin, P. Wenk, R.~M. Westervelt, and R. Winkler for
collaborations on topics related to this article.
Moreover, I am grateful to J. Fabian, S.~D. Ganichev, J. Nitta, J. Orenstein, 
K. Richter, G. Salis, and D. Weiss for the permission to reproduce here 
figures of their publications. Thanks for useful comments on the manuscript 
go to P. Wenk.

This work was supported by Deutsche Forschungsgemeinschaft via SFB 689.

\bibliography{pssrev}% Produces the bibliography via BibTeX.

\end{document}